\documentstyle[psfig]{elsart}
\def\>{\rangle}\def\<{\langle}\def\^{\hat}
\def\defi{\stackrel{\mbox{\tiny def}}{=}}

\def\Tr{\mbox{Tr}}
\def\+{\dagger}

\begin{document} 

\begin{frontmatter}
\title{{Orthogonality relations in Quantum Tomography}}
\author{G. M. D'Ariano, L. Maccone,\thanksref{CA}}
\author{ and M.G.A. Paris}
\address{Theoretical Quantum Optics Group,\\ INFM and
Dipartimento di Fisica `Alessandro Volta' \\ Universit\`a di Pavia,
via Bassi 6, I-27100 Pavia, ITALY}

\begin{abstract}
{Quantum estimation of the operators of a system is investigated by
analyzing its Liouville space of operators. In this way it is possible
to easily derive some general characterization for the sets of
observables ({\it i.e.} the possible {\it quorums}) that are measured
for the quantum estimation.  In particular we analyze the
reconstruction of operators of spin systems.}
\end{abstract}
\thanks[CA]{Corresponding author. E-mail address:
{\tt maccone@unipv.it}}
\end{frontmatter}

\section{Introduction}
Two fundamental restrictions limit the possibility of devising a state
reconstruction method. On one hand, the quantum complementarity
principle does not allow to recover the quantum state from
measurements on a single system, unless we have some prior information
on it. On the other hand, the no cloning theorem ensures that it is
not possible to make exact copies of a quantum system, without having
prior knowledge of its state.  Hence, the {only possibility} for
devising a state reconstruction procedure is to provide a measuring
strategy that employs numerous identical (although unknown) copies of
the system, so that different measurements may be performed on each of
the copies.\par The problem of state estimation resorts essentially to
estimating arbitrary operators of a quantum system by using the result
of measurements of a set of observables. If this set of observables is
sufficient to give full knowledge of the system state, then we define
it a quorum. Notice that, in general, a system may allow various,
different quorums. Quantum tomography was born {\cite{tomog}} as a
state reconstruction technique in the optical domain, and has recently
been extended {\cite{grouptom}} to a vast class of systems. By
extension, we now denote as ``Quantum Tomography'' all unbiased
quantum state reconstruction procedures, {\it i.e.} those procedures
which are affected only by statistical errors that can be made
arbitrarily small by increasing the number of measurements.
Tomography makes use of the results of the quorum measurements in
order to reconstruct the expectation value of arbitrary operators
(even not observables) acting on the system Hilbert space.\par The
purpose of this work is to present in a formally familiar manner
(employing the Dirac notation also on operator space) a constructive
method to derive tomographic formulas for quantum systems, at least
for finite dimensional Hilbert spaces. This is achieved by giving
conditions to build quorums and to check whether a given set of
operators is a quorum.  In this way, we obtain an extension of the
recently proposed group tomography {\cite{grouptom}}, where similar
conditions were derived for systems with an underlying group
structure.
\par
In Sect. {\ref{general}} we give the definitions and the conditions to
identify a quorum of operators by analyzing the space of operators of
a system as a linear vector space. We derive a constructive
algorithmic procedure to obtain tomographic formulas in the case of
finite quorums. In Sect. {\ref{spin}} we give some examples of
applications of the presented method in the domain of spin systems,
where various different quorums are available
{\cite{grouptom,weigert}}.

\section{General estimation}\label{general}
Consider the set of system operators, {\it i.e.} the Liouville space
${\cal L}({\cal H})$. If we initially restrict ourselves to
Hilbert-Schmidt operators in ${\cal L}({\cal H})$, then this set is
itself a Hilbert space of operators, with the scalar product
\begin{eqnarray}
\langle\hat A|\hat B\rangle\defi \hbox{Tr}\left[ \hat A^\dagger\:
\hat B\right] 
\;\label{scalprod}.
\end{eqnarray}
It is then possible to employ all the properties of linear vector algebra,
and to use the Dirac notation, by using the following definitions for
bra and ket vectors:
\begin{eqnarray}
\hat O&\longrightarrow&|\hat O\>\nonumber\\
\Tr[\bullet\hat
O^\dagger]&\longrightarrow&\<\hat O|\bullet\;\label{brakets}.
\end{eqnarray}
In this vision, quantum tomography consists of expressing the operator
$\^A$ we want to evaluate
as an expansion on the observables of the quorum as
\begin{eqnarray}
|\hat A\>=\int_{\cal X}dx\;|\hat C(x)\>\<\hat B(x)|\hat A\>\;
\;\label{tomog},
\end{eqnarray}
where $|\hat A\>$ is a generic operator in ${\cal L}({\cal H})$,
$|\hat C(x)\>$ (with $x\in{\cal X}$) is the set of quorum observables
($C(x)$ is a generally complex function of a selfadjoint operator,
hence it is observable in this sense), and the set $\<\^B(x)|$ is the
dual of the quorum.  In ordinary notation, Eq. (\ref{brakets}) is the
tomography identity, {\it i.e.} \begin{eqnarray}
\^A=\int_{\cal X}dx\;\Tr\left[\^B(x)^\dagger\^A\right]\^C(x)
\;\label{ordintom}
\end{eqnarray}
Notice that the extension of the theory to non-normalizable vectors in
the operator Hilbert space is immediate: one only has to require the
existence of the trace of Eq. (\ref{ordintom}). If, for example, $\^A$
is a trace--class operator, then we do not need to require $\^B(x)$ to
be of Hilbert-Schmidt class, since it is sufficient to require
$\^B(x)$ bounded.  Through Eq. (\ref{tomog}), the tomographic
reconstruction procedure is immediately obtained. In fact, by
measuring the observables $|\hat C(x)\>$ of the quorum, we
can(\footnote{Eq. (\ref{recoproc}) is obtained by taking the
expectation value of both members of Eq. (\ref{tomog}) and by
calculating the expectation value trace using the eigenvectors of the
quorum observables $|\^C(x)\>$.}) express the mean value of any
operator $\<\hat A\>$ in terms of the eigenvalues of $|\hat C(x)\>$ as
\begin{eqnarray}
\<\^A\>=
\int_{\cal X}dx\;\sum_mp(m,x)\;
\lambda^{(x)}_m\;\Tr[\^B^\dagger(x)\^A]\;,
\;\label{recoproc}
\end{eqnarray}
where $p(m,x)$ is the probability of obtaining the eigenvalue
$\lambda^{(x)}_m$ when measuring the quorum observable $\^C(x)$.\par

Since we want Eq. (\ref{tomog}) to be valid for a {generic} operator
$|\hat A\>$ in ${\cal L}({\cal H})$ [or also in a subspace of ${\cal
L}({\cal H})$], then we must require that the $|\hat C(x)\>$
constitute a spanning set for the operator (sub)space, with the set of
$\<\hat B(x)|$ acting as its dual. A spanning set is a generalized
basis for a vector space: it is a complete set of vectors but it is
not, in general, composed of linearly independent (or normalized)
vectors.  Define {dual} $\<\^B(x)|$ of the set $|\hat C(x)\>$ as the
set constructed so to have
\begin{eqnarray}
\<\hat B(x)|\hat
C(x')\>=\Tr[\hat B^\dagger(x)\hat C(x')]=\delta(x,x')\;\forall
x,x'\in{\cal X},\;\label{dualdef}
\end{eqnarray}
where $\delta(x,x')$ is a reproducing kernel for $\<\^B(x)|$, {\it
i.e.}
\begin{eqnarray}
\int_{\cal X}dx\;\delta(x,x')\<\^B(x)|=\<\^B(x')|\;.\label{delta1}
\end{eqnarray}
Since $|\^C(x)\>$ is a complete set, $\delta(x,x')$ is a reproducing
kernel also for this set, {\it i.e.}
\begin{eqnarray}
\int_{\cal X}dx\;\delta(x,x')|\^C(x)\>=|\^C(x')\>\;.\;\label{delta2}
\end{eqnarray}

From linear vector algebra we obtain the following four equivalent
definitions of spanning set:\\ A set of vectors $|\^C(x)\>$ (with dual
$\<\^B(x)|$) is a spanning set $\Leftrightarrow$
\begin{enumerate}
\item[i)]  $\displaystyle\forall |\^A\>\in{\cal L}({\cal
H}),\;|\^A\>=\int_{\cal X}dx\; |\^C(x)\>\<\^B(x)|\^A\>$, {\it i.e.}
the tomographic identity, namely Eq. (\ref{tomog}).
\item[ii)] $|\^C_n\>$ is complete, {\it i.e.} {(no
nonzero element is orthogonal to $|\^C(x)\>$ $\forall x$)}:
\begin{eqnarray}
\<\^A|\^C(x)\>=\<\^B(x)|\^A\>=0\;\forall x\in{\cal
X}\Rightarrow|\^A\>=0
\;\label{completezza1}.
\end{eqnarray}
\item[iii)] the following operatorial identity resolution applies,
\begin{eqnarray}
\int_{\cal X}dx\;|\^C(x)\>\<\^B(x)|=\^{\^{1}}\;,\label{opidres}
\end{eqnarray}
where $\hat{\hat{1}}$ is the identity super-operator, { namely the
operator acting on operators such that
$\hat{\hat{1}}[\^A]=\^A\;\forall\^A\in{\cal L}({\cal
H})$.}\label{condtre}
\item[iv)]
$\displaystyle\int_{\cal
X}dx\;\<\^A|\^C(x)\>\<\^B(x)|\^A\>=\parallel\^A\parallel^2\;\defi
{\Tr[\^A^\dagger\^A]}\ \ \forall|\^A\>\in{\cal L}({\cal H}).$
\end{enumerate}
In the usual notation, these equivalent definitions write
as {\cite{matteo}}:\begin{enumerate}
\item[i)] $\displaystyle\^A=\int_{\cal X}dx\;\Tr[\^B^\dagger(x)\^A]\^C(x)$.
\item[ii)] $\Tr[\^B^\dagger(x)\^A]=\Tr[\^A^\dagger\^C(x)]=0\ \forall
x\in{\cal X}\Rightarrow\^A=0$.
\item[iii)] $\displaystyle\int_{\cal X}dx\;
\<i|\^C(x)|j\>\<k|\^B^\dagger(x)|l\>=\delta_{il}\delta_{jk}$, where
$\{|n\>\}$ is a basis for the system Hilbert space $\cal H$.
\item[iv)]
$\displaystyle\int_{\cal
X}dx\;\Tr[\^A^\dagger\^C(x)]\Tr[\^B^\dagger(x)\^A]=
{\Tr[\^A^\dagger\^A]}\ \ \forall\^A\in{\cal L}({\cal H})$.
\end{enumerate}

In order to obtain the dual set $\<\^B(x)|$ starting from a given set
$|\^C(x)\>$, one in general has to solve the operatorial equation
(\ref{dualdef}) that defines the quorum. For finite quorums, this
resorts to a matrix inversion. An alternative procedure is now
proposed. It derives from the Gram--Schmidt orthogonalization method
{\cite{richt}}, which allows to derive a basis starting from a
complete set of vectors.\label{gsproc} Namely, one obtains a basis
$|y_k\rangle$, given the complete set $|C_k\rangle$ (assume for
simplicity that all $|C_k\rangle$ are non-zero and that in
$\{|C_k\rangle\}$ there are no couples of proportional vectors),
recursively defined as
\begin{eqnarray}
\left\{\matrix{|y_0\rangle\doteq \frac1{N_0}\;|C_0\rangle\nonumber\\
|y_{k}\rangle\doteq
\frac1{N_k}\;\left(|C_{k}\rangle-\sum_{j=0}^{k-1}|y_j\rangle\langle 
y_j|C_{k}\rangle\right)} \right.
\;\label{gramschmidt},
\end{eqnarray}
where $N_0\doteq\parallel|C_0\rangle\parallel$ and $N_k\doteq\parallel
|C_{k}\rangle-\sum_{j=0}^{k-1}|y_j\rangle\langle
y_j|C_{k}\rangle\parallel$.  Notice that in the recursion
(\ref{gramschmidt}) one must take care of eliminating all the vectors
$|C_k\>$ which are a linear combination of the $|y_j\>$ with $j<k$.

Write the identity resolution for the basis obtained with procedure
(\ref{gramschmidt}), {\it i.e.}
\begin{eqnarray}
\hat 1&=&\sum_{k=0}{|y_k\rangle}\langle y_k|\equiv\nonumber\\
&&\frac{|C_0\rangle}{N_0}\langle
y_0|+\sum_{k=1}\frac
1{N_k}\left(|C_k\rangle-\sum_{j=0}^{k-1}{|y_j\rangle}\langle
y_j|C_k\rangle\right)\langle y_k| \label{gsidres}.
\end{eqnarray}
By using repeatedly Eq. (\ref{gramschmidt}) (expressing $|y_j\rangle$
of Eq. (\ref{gsidres}) in terms of the $|C_n\rangle$s) and by
reorganizing the terms in the sums, we can find the dual set $\langle
B_n|$ as
\begin{eqnarray}
&&\langle B_0|=\frac {\langle y_0|}{N_0}-\frac{\langle
y_0|C_1\rangle\langle y_1|}{N_0N_1} +\left(-\frac{\langle
y_0|C_2\rangle}{N_0N_2}+\frac{\langle
y_0|C_1\rangle\langle y_1|C_2\rangle}{N_0N_1N_2}\right)\langle
y_2|+\cdots \nonumber\\
&&\langle B_1|=\frac{\langle y_1|}{N_1} -\frac{\langle
y_1|C_2\rangle\langle y_2|}{N_1N_2}+\left(-\frac{\langle
y_1|C_3\rangle}{N_1N_3}+\frac{\langle
y_1|C_2\rangle\langle y_2|C_3\rangle}{N_1N_2N_3}\right)\langle
y_3|+\cdots\nonumber\\ &&\cdots 
\;\label{gscob}
\end{eqnarray}
Eq. (\ref{gsidres}) guarantees that it is possible to write
\begin{eqnarray}
\hat 1=\sum_n|C_n\>\<B_n|
\;\label{idresnuova},
\end{eqnarray}
which is just the definition iii [{\it i.e.} Eq. (\ref{opidres})] of
spanning set.

Summarizing, we described a method for deriving tomographic formulas
for arbitrary systems. One must start from a set of operators $\^C_n$
he would like to use as a quorum, and verify that such a set is
complete, {\it i.e.} that no nonzero element of ${\cal L}({\cal H})$
is orthogonal to all $\^C_n$: \begin{eqnarray}
\<\^A|\^C_n\>=\Tr[\^A^\dagger\^C_n]=0\;\forall n\Rightarrow|\^A\>=0
\;\label{completezza}.
\end{eqnarray}
If the set is finite, then one can employ the orthogonalization
procedure outlined previously to derive the dual set. If the set is
infinite discrete or continuous, then one can only resort to finding
appropriate solutions for Eq. (\ref{dualdef}). Once the dual is known,
the tomographic identity (\ref{tomog}) can be written explicitly.  The
reconstruction procedure, in terms of the probabilities of
measurements of quorum observables, follows straightforwardly and
yields Eq. (\ref{recoproc}), which allows to obtain arbitrary operator
expectation values in terms of quorum outcome probabilities. Of
course, one may think of similar procedures based on different
orthogonalization algorithms.\par Since no hypotheses were made on the
structure of the system Hilbert space, the theory presented in this
section is valid for any quantum system. In the following section we
will give some example applications.

\section{Example of application: Spin Tomography}\label{spin}
Here we show an application of the theory presented in the previous
section by rederiving the spin tomography {\cite{grouptom,weigert}},
where various different quorums may be employed.

The {simplest} possible example is a {spin $s=\frac 12$} system. In
this case we expect that the Pauli matrix and the identity constitute
a quorum (since any $2\times 2$ matrix can be written on such a
basis). Take the quorum {$\displaystyle{\cal
Q}\defi\{\^\sigma_x,\^\sigma_y,\^\sigma_z,\^1\}$:} it is immediate to
verify that it is complete. Since the quorum operators are orthogonal,
{\it i.e.}
$\^\sigma_\alpha\cdot\^\sigma_{\alpha'}=\^1\delta_{\alpha\alpha'}$
($\alpha,\alpha'=x,y,z$), using the Gram-Schmidt procedure it is
immediate to obtain the dual set as {${\cal C}=\{\frac 12\^\sigma_x,
\frac 12\^\sigma_y, \frac 12\^\sigma_z, \frac 12\^1\}$.} The
expansion (\ref{tomog}) of a matrix $\^A$ is, thus
\begin{eqnarray}
|\^A\>=\frac
12\left[\sum_{\alpha=x,y,z}|\^\sigma_\alpha\>
\<\^\sigma^\dagger_\alpha|\^A\>
+ |\^1\>\<\^1|\^A\>\right]\;,\;\label{mat2p2}
\end{eqnarray}
which {immediately} yields the reconstruction
procedure\begin{eqnarray}
\langle\hat A\rangle=\sum_{m=-\frac 12}^{\frac
12}\;\sum_{\alpha=x,y,z} p(m,\vec n_\alpha)\;m\; \hbox{Tr}
\left[\hat A\hat\sigma_\alpha\right]
+\frac 12 \: \hbox{Tr}
\left[\hat A\right]
\;,\end{eqnarray}
where $p(m,\vec n_\alpha)$ is the probability to obtain the eigenvalue
$m=\pm\frac 12$ while measuring $\vec S\cdot\vec n_\alpha$.  This
equation allows the {reconstruction of the expectation value of any
spin $s=\frac 12$ operator $\^A$ from the measurement of the spin in
the $x,y,z$ directions.}
\vskip 1\baselineskip\par

For {an arbitrary spin $s$}, a {possible quorum} is given by the spin
component in all directions, {\it i.e.}  the {observable $\vec
S\cdot\vec n$} ($\vec S$ being the spin operator and $\vec n$ a vector
on the unit sphere). In order to find the dual $\<\hat B|$, consider
the exponential of the quorum, {\it i.e.} $\hat D(\psi,\vec
n)=\exp(i\psi\vec S\cdot\vec n)$, which satisfies definition iii [{\it
i.e.} Eq. (\ref{opidres})] of spanning set. In fact, $\hat D(\psi,\vec
n)$ constitutes a unitary irreducible representation of the group
SU(2).  The orthogonality relation between the matrix elements of the
group representation $D(g)$ of dimension $d$ writes as {\cite{murna}}
\begin{eqnarray}
\int_R dg\;{D}_{jr}(g){D}^\dagger_{tk}(g)=\frac
Vd\delta_{jk}\delta_{tr}\;\label{ortmurna},
\end{eqnarray} where $dg$ is the group Haar invariant
measure, and $V=\int_Rdg$. For SU(2), with the $2s+1$ dimension
unitary irreducible representation $\hat D(\psi,\vec n)$, Haar's
invariant measure is $\sin^2\frac\psi2\sin\vartheta\; d\vartheta\;
d\varphi\; d\psi$, and $V={4\pi^2}$. Thus, the orthogonality relations
in this case are given by \begin{eqnarray}
\frac{2s+1}{4\pi^2}\int_\Omega d\vec n
\int_0^{2\pi}d\psi\;\sin^2\frac\psi2 \langle j|e^{i\psi\vec n\cdot\vec
S}|r\rangle\langle t|e^{-i\psi\vec n\cdot\vec
S}|k\rangle=\delta_{jk}\delta_{tr}\;,\;\label{sporth}
\end{eqnarray} 
which is the the spanning set definition iii for the set of operators
$|\^D\>=\^D$, with dual
$\<\^D^\dagger|\bullet=\Tr[\^D^\dagger\bullet]$.\par Then, it is
possible to write the spin tomography identity as
\begin{eqnarray} 
\hat A=\frac{2s+1}{4\pi^2}\int_\Omega d\vec n
\int_0^{2\pi}d\psi\;\sin^2\frac\psi2 \: \hbox{Tr} \left[\hat A\hat
D^\dagger(\psi,\vec n)\right]\hat D(\psi,\vec n)
\;\label{spintom},
\end{eqnarray}
from which the following reconstruction procedure is derived
\begin{eqnarray}
\langle\hat A\rangle=\frac{2s+1}{4\pi^2}\sum_{m=-s}^s\int_{\Omega}d\vec
n\;p(m,\vec n)\int_0^{2\pi}d\psi\;\sin^2
\frac\psi2 \: \hbox{Tr}
\left[\hat A\; e^{-i\psi\left(\vec S\cdot\vec n-m\right)}\right]
\;\label{spinre},\end{eqnarray}
where $p(m,\vec n)$ is the probability of obtaining $m$ as the
measurement result of $\vec S\cdot\vec n$. This equation allows the
reconstruction of arbitrary spin $s$ expectation values $\<\^A\>$,
from spin measurements in all directions $\vec n$.

\par
\begin{figure}
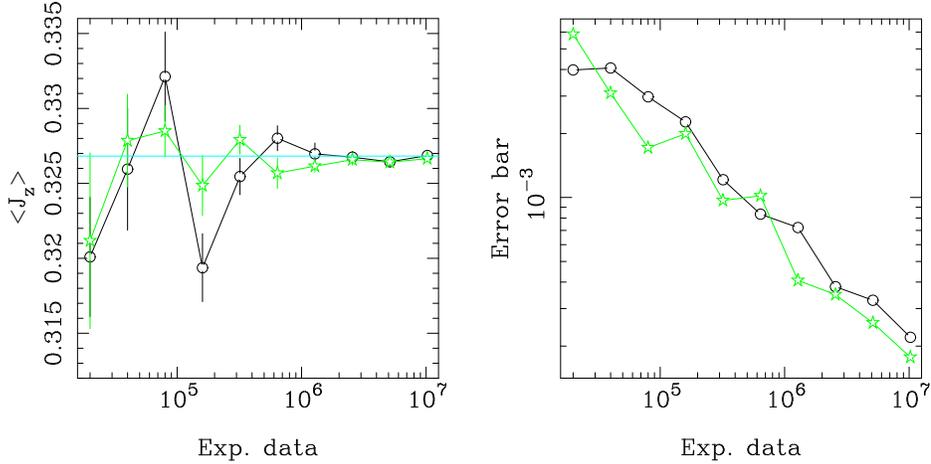

\begin{center}
\begin{tabular}{cc}
\psfig{file=contvsdisc1.ps,width=6cm}
&
\psfig{file=contvsdisc2.ps,width=6cm}
\end{tabular}
\end{center}
\caption{%
{Monte Carlo comparison between continuous and discrete tomography for
a spin $s=\frac 12$ system.} {Continuous tomography uses
$\^D(\psi,\vec n)$ as quorum, while discrete tomography uses the
quorum ${\cal
Q}\defi\{\^\sigma_x,\^\sigma_y,\^\sigma_z,\^1\}$. {Left:} Convergence
of the mean value of $\langle s_z\rangle$ for a coherent $\alpha=2$
spin state for increasing number of experimental data (the theoretical
value is given by the horizontal line).  The circles $\circ$ refer to
continuous, the stars {$\star$} to discrete tomography. {Right:} Plot
of the statistical error bars of the graphs on the left {\it vs}
experimental data. The error bars are obtained by dividing the
experimental data into $20$ statistical blocks. Notice that the two
tomographic procedures are essentially equivalent.}}
\label{f:discr}\end{figure}

 
Numerical simulations show that the two preceding quorums are (for
spin $s=\frac 12$) equivalent, namely the same number of experimental
measurement data yield the same results and the same statistical error
bars, apart from statistical fluctuations. In Fig. {\ref{f:discr}} a
Monte Carlo comparison of the two spin reconstruction strategies based
on the two different quorums is given. Both reconstructions are
applied to a coherent spin state, defined as
$|\alpha\>\defi\exp(\alpha S_+-\alpha^*S_-)|-s\>$, where $S_+,S_-$ are
the spin lowering and raising operators and $|-s\>$ is the eigenvector
of $S_z$ relative to the minimum eigenvalue. 

\vskip 1\baselineskip\par
{Weigert} has shown {\cite{weigert}} that another spin $s$ quorum can
be obtained by taking {$N_s\defi(2s+1)^2$ arbitrary}(\footnote{
Actually the choice of the directions is not completely arbitrary, but
``almost'' {\cite{weigert}} any choice yields a complete set of
operators in ${\cal L}({\cal H})$.})  directions $\vec n_k$ and
measuring the observables $\displaystyle\^{\cal Q}_k\defi|\vec
n_k\>\<\vec n_k|$, which are the projectors for the eigenspace
relative to the maximum eigenvalue $s$ of the observables $\vec
S\cdot\vec n_k$. We {define a dual} $\<\^{\cal Q}_k|$ for the
$|\^{\cal Q}_k\>$ by requiring
\begin{eqnarray}
\<\^{\cal Q}_k|\^{\cal Q}_{k'}\>=\delta_{kk'}\;,\label{ortw}
\end{eqnarray}
{\it i.e.} Eq. (11) of {\cite{weigert}}, 
which is just the
dual set definition (\ref{dualdef}).  Condition (\ref{ortw}) together
with the completeness of the chosen quorum, guarantee that $|\^{\cal
Q}_k\>$ (with dual $\<\^{\cal Q}_k|$) is a spanning set for ${\cal
L}({\cal H})$, thus allowing the tomographic identity \begin{eqnarray}
|\^A\>=\sum_{k=1}^{N_s}|\^{\cal Q}_k\>\<\^{\cal Q}_k|\^A\>
\label{weigt1}\;,\end{eqnarray}{\it i.e.} {
(using the notation of {\cite{weigert}})}\begin{eqnarray}
\^A=\sum_{k=1}^{N_s}\Tr[\^A\^{\cal Q}^k]\^{\cal Q}_k
\;,\label{weigtom}
\end{eqnarray}
where $\^{\cal Q}^k$ is the dual operator of $\^{\cal Q}_k$. The
{explicit form} of the dual set $\^{\cal Q}^k$ can be derived by a
matrix inversion starting from Eq. (\ref{ortw}) or by the
Gram--Schmidt based procedure method given on page
\pageref{gsproc}. The {reconstruction procedure} is, in this case,
\begin{eqnarray}
\langle\hat A\rangle=s\sum_{k=1}^{N_s} p(s,\vec n_k)\: \hbox{Tr}
\left[\hat A\^{\cal Q}^k\right]
\;\label{recweig},
\end{eqnarray}
where $p(s,\vec n_k)$ is the probability of obtaining the maximum
eigenvalue $s$, when measuring $\vec S\cdot\vec n_k$. {This allows the
reconstruction of arbitrary spin operators $\^A$ from measurements of
the spin along $N_s$ fixed directions.}

\section{Conclusions}
Recent {Group Tomography \cite{grouptom}} gives a general framework
that allows to derive all the state reconstruction procedures that
employ quorums which exhibit a group symmetry.  Here we extended these
results to generic state reconstruction procedures.  In fact, we have
seen how it is possible to give a {characterization of tomographic
formulas in terms of linear vector algebra} on the vectors of the
Liouville space of the system.\par A {constructive method to derive
new tomographic formulas} has been proposed starting from the
Gram--Schmidt orthogonalization procedure. {At least in principle, it
allows to calculate the quorum dual for the quantum systems that allow
a discrete quorum.} We have given some {examples of the method in the
spin domain,} by re-obtaining all the known spin tomographies using
linear vector algebra arguments.  For the sake of illustrating the
method, we limited our analysis to the description of spin systems,
but all known tomographies can be analyzed in this framework
{\cite{matteo}}.  Moreover, one may expect to employ the presented
procedures to uncover {new tomographies} for quantum systems for which
state reconstruction procedures are not presently known.

\ack This work has been partially supported by INFM
through project PAIS-1999-TWIN.

\end{document}